\documentclass[aps,pre,twocolumn,groupedaddress]{revtex4-1}

\usepackage{amsmath}
\usepackage{amssymb}
\usepackage{graphicx}

\begin{document}

\title{Enhanced ability of information gathering may intensify disagreement among groups}

\author{Hiroki Sayama}
\email{sayama@binghamton.edu}
\homepage{http://bingweb.binghamton.edu/~sayama/}
\affiliation{Center for Collective Dynamics of Complex Systems,
Binghamton University, Binghamton, NY 13902-6000, USA}
\affiliation{Max Planck Institute for the Physics of Complex Systems,
01187 Dresden, Germany}
\affiliation{Waseda Innovation Lab, Waseda University,
Shinjuku, Tokyo 169-8050, Japan}

\date{\today}

\begin{abstract}
Today's society faces widening disagreement and conflicts among
constituents with incompatible views. Escalated views and opinions are
seen not only in radical ideology or extremism but also in many other
scenes of our everyday life. Here we show that widening disagreement
among groups may be linked to the advancement of information
communication technology, by analyzing a mathematical model of
population dynamics in a continuous opinion space. We adopted the
interaction kernel approach to model enhancement of people's
information gathering ability and introduced a generalized non-local
gradient as individuals' perception kernel. We found that the
characteristic distance between population peaks becomes greater as
the wider range of opinions becomes available to individuals or the
greater attention is attracted to opinions distant from theirs. These
findings may provide a possible explanation for why disagreement is
growing in today's increasingly interconnected society, without
attributing its cause only to specific individuals or events.
\end{abstract}

\maketitle

\section{Introduction}

Today's society faces many urgent critical challenges. One of such
challenges is addressing the widening disagreement and conflicts among
different social constituent groups with incompatible views on
politics, economy, international relationships, religions, cultures,
lifestyle, and other aspects of our life.

Studies on this challenge often focus on how escalated views and
opinions emerge in society \cite{chuang2019mathematical}. Typical
approaches in this area include detection of extremism in online media
\cite{ferrara2016predicting,badawy2018rise,manrique2018generalized}
and modeling contagious processes of extremism through social networks
\cite{berger2015metronome,ferrara2017contagion}. While highly relevant
to and valuable for national security concerns, these approaches
necessarily impose an asymmetric point of view to consider one side as
the cause of the problem (``us'' being normal vs.\ ``them'' being
abnormal), making it difficult to obtain a more system-oriented
understanding of how such conflicts may arise and widen spontaneously
at a global societal scale.

Escalated views and opinions are seen not only in radical ideology or
extremism, but also in many other scenes of our everyday life
(typically in a milder form), such as political conversations in
social media \cite{conover2011political,morales2015measuring},
healthcare choices (e.g., anti-vax movement) \cite{kata2012anti,johnson2019health}, and dietary preferences \cite{cole2011vegaphobia,reilly2016gluten}, to name a few. The widening disagreement
among those who have escalated views is becoming more prevalent than
before on a variety of subjects. Part of the cause is often suspected
to be the recent advances of information communication technology
(e.g., web media, social media, smart phones, and other forms of
high-speed, high-volume, personalized communication)\cite{kata2012anti,prior2013media,bakshy2015exposure,bail2018exposure} that
continuously increase information gathering intensity and enhance
users’ ability to choose their preferred information sources (with some caveats \cite{boxell2017greater}). In this view, widening disagreement may be modeled and
understood as a spontaneous pattern formation process in which
people's information gathering ability plays a key role as the control
parameter. The present study explores this view through mathematical
modeling and analysis of opinion dynamics.

In this study, we combine the opinion
dynamics with spatial models studied in mathematical
biology. Specifically, we describe the dynamics of popularities of a
wide range of opinions in partial differential equations (PDEs),
inspired by models of diffusion and migration of biological organisms
\cite{edelstein2005mathematical,sayama2015introduction}. In
particular, we adopt a model of auto-aggregation
\cite{keller1970initiation,horstmann20031970,boi2000modeling,hillen2009user}
in which organisms aggregate together through a hill-climbing
migration behavior on a terrain of signals. In our case, we consider
people migrating in a space of opinions, and their migration is driven
by the gradient of the opinion popularity itself.

We also propose a principled way to generalize local gradient into
non-local perceived gradient based on the interaction kernel approach
used in physics, applied mathematics, and mathematical biology. This
allows us to model the enhancement of people's information gathering
ability, which would not be captured by simply using a conventional
local gradient at a single point in the opinion space. It also allows
for parametrization of two distinct aspects of non-local perception:
the breadth of information gathering and the level of selective
attention paid to distant opinions, the latter of which we hypothesize
to play a particularly significant role in social opinion drift
\cite{sayama2015social,sayama2016going}. The model and the results of
our analytical and numerical investigations are reported below.

\section{Model}

\begin{figure*}[htbp]
\centering
\includegraphics[width=\textwidth]{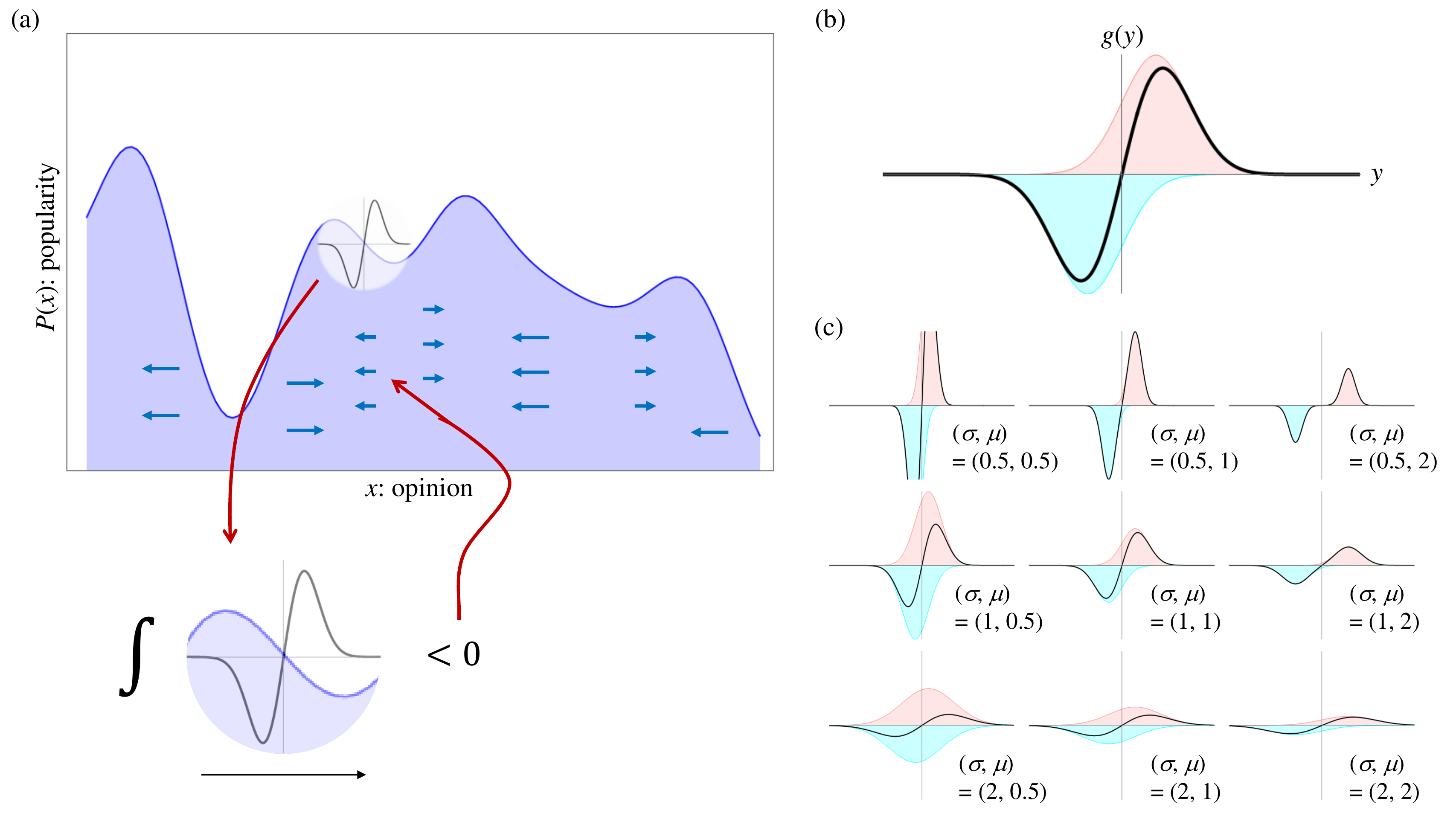}
\caption{Schematic overview of the mathematical model proposed in this
  study.  (a) Popularities of opinions represented as
  distribution $P(x,t)$ on opinion $x$. $P(x,t)$ follows
  diffusion-migration dynamics described in Eq.~(\ref{eq:model}),
  which may show, under typical parameter settings, aggregation
  behaviors as illustrated by small horizontal arrows in this
  figure. Directions of migrations are determined by a perceived
  gradient $G(P)$ defined as a cross-correlation between $P(x,t)$ and
  a perception kernel $g(y)$ at each location $x$ (Eq.~(\ref{eq:GP}),
  also depicted at the bottom of this panel).  (b) Structure of
  the perception kernel $g(y)$ (Eq.~(\ref{eq:kernel})). The black
  solid curve shows the shape of $g(y)$, which is the sum of two
  Gaussian distributions with opposite signs (pink and cyan), one
  placed on the right and another on the left.  (c) Variations
  of shapes of $g(y)$ as $\sigma$ and $\mu$ are varied. Having sharp
  peaks near the center in $g(y)$ (top left) makes $G(P)$ close to a
  conventional local derivative, corresponding to the case in which
  individuals' attention is limited only to opinions of similarly
  minded others. Having broader and/or distant peaks means enhanced
  information gathering ability, covering a wider range of opinions
  and/or paying greater attention to distant opinions, respectively.}
\label{fig:overview}
\end{figure*}

Our mathematical model represents the dynamics of popularities of
opinions using a population distribution function $P(x,t)$ for opinion
$x$ and time $t \geq 0$ (Fig.~\ref{fig:overview}a, top). $P(x,t)$ is
the number (in an arbitrary unit) of people whose current opinion is
$x$ at time $t$. We assume the population never grows or decays, so
the only changes possible in this model are due to diffusion and
migration. Diffusion represents random fluctuations of people's
opinions, while migration represents directed, active change of
people's opinions caused by social influence. More specifically, we
adopt a widely used assumption \cite{degroot1974reaching,friedkin1990social,banisch2019opinion} that people will
more likely adopt opinions that are more supported by others. We also
assume the homophily principle \cite{mcpherson2001birds,kossinets2009origins,bakshy2015exposure} in people's information
gathering behavior, which implies that they perceive information only
from a vicinity of their own opinion in the opinion space. The last
two assumptions simplify the migration process into a simple
hill-climbing behavior following the gradient of $P(x,t)$. Note that the opinion space modeled here is different from physical space within which individuals exchange opinions (e.g., social networks). Such social structure is not modeled explicitly in this study. 

Both diffusion and migration can be modeled using the transport equation framework \cite{sayama2015introduction,hillen2009user,di2013measure}. The model equation we use in this study to describe the dynamics of $P(x, t)$ is
\begin{align}
\frac{\partial P}{\partial t} &= d \nabla^2 P - c \nabla \cdot \left( P G(P)\right) ,\label{eq:model}
\end{align}
where $d \nabla^2 P$ is the diffusion term and $-c \nabla \cdot (P G(P))$ is the migration term. $G(P)$ is the {\em perceived} gradient of popularity distribution, defined as 
\begin{align}
G(P) &= \int_{-\infty}^\infty P(x+y, t) g(y) dy ,\label{eq:GP}\\
g(y) &= \frac{1}{2 \mu}\cdot\frac{1}{\sqrt{2\pi} \sigma}
\left( e^{-\frac{1}{2} \left(\frac{y-\mu}{\sigma}\right)^2} 
- e^{-\frac{1}{2} \left(\frac{y+\mu}{\sigma}\right)^2} \right). \label{eq:kernel}
\end{align}
Eq.\ (\ref{eq:GP}) shows that the perceived gradient is defined as a
cross-correlation between $P$ and a perception kernel $g$ given in
Eq.\ (\ref{eq:kernel}) (Fig.~\ref{fig:overview}a, bottom), based on
the interaction kernel approach commonly used in physics, applied
mathematics, and mathematical biology
\cite{boi2000modeling,di2013measure,mogilner1999non}. The perception
kernel $g$ describes how people assign weights (attentions) to nearby
opinions' popularities when they assess the gradient. In this study,
we define the perception kernel as a combination of two Gaussian
distributions whose width is determined by $\sigma$ and whose means
are separated by $2\mu$ across the origin, one positive on the right
hand side and one negative on the left hand side, to capture the
difference of popularity levels between the two sides
(Fig.~\ref{fig:overview}b).

Note that $G(P)$ converges to a simple derivative $P'(x)$ in the limit
of $\mu \to 0^+$ and $\sigma \to 0^+$, i.e., when $g$ is made of two
Dirac's delta functions positioned right next to the origin with
opposite signs (see Appendix \ref{convergence}). This indicates that $G(P)$
can be considered a mathematically valid non-local generalization of a
spatial derivative.

This generalization of the gradient enables us to explore different
shapes of the perception kernel by varying $\sigma$ and $\mu$
(Fig.~\ref{fig:overview}c) and study their effects on opinion
dynamics, which would not have been possible if only local gradient
were used. For example, increasing $\sigma$ (Fig.~\ref{fig:overview}c,
bottom left) represents a situation in which people can gather
information from a broader range of opinions. Meanwhile, increasing
$\mu$ (Fig.~\ref{fig:overview}c, top right) represents a situation in
which people tend to pay great attention particularly to distant,
extreme opinions, which may correspond to sensationalism widely
practiced in various media today.

\section{Results}

\subsection{Stability analysis}

We first conduct linear stability analysis of Eq.\ (\ref{eq:model}) to
find the conditions under which homogeneous population distributions
are unstable and thus heterogeneous patterns will form.  We begin the
analysis by replacing the spatio-temporal function $P(x,t)$ with a
constant homogeneous population level $P_h$ plus a sinusoidal spatial
perturbation with temporally changing small amplitude $\Delta P(t)$
\cite{sayama2015introduction}, as follows:
\begin{align}
P(x, t) &\to P_h + \Delta P(t) \sin (\omega x + \phi)
\end{align}
This replacement allows for linearization of Eq.\ (\ref{eq:model})
into the following non-spatial linear dynamical equation of $\Delta P$
(see Appendix \ref{stability-analysis} for details):
\begin{align}
\frac{d \Delta P}{d t} &= \left( - d \omega^2  + c \omega P_h \int_{-\infty}^\infty \sin(\omega y) g(y) dy \right) \Delta P 
\end{align}
Therefore, with $\displaystyle Q(\omega) = \int_{-\infty}^\infty \frac{\sin(\omega y)}{\omega} g(y) dy$, if
\begin{align}
Q(\omega) &> \frac{d}{c P_h} \label{eq:condition}
\end{align}
for $\omega > 0$, then the homogeneous population distribution is
unstable and a non-homogeneous pattern (i.e., islands of popular
opinions = distinct groups) will form in the opinion space. This
result already tells us that groups are more likely to form if (1)
diffusion is weaker, (2) active migration is stronger, and/or (3) the
population level is greater. These findings are consistent with
results obtained for other similar auto-aggregation models
\cite{edelstein2005mathematical,sayama2015introduction,keller1970initiation,horstmann20031970}.

We also note that the range of $Q(\omega)$ is bounded to $[-1, 1]$ and
$Q(\omega)$ approaches its maximum 1 as $\omega \to 0$ regardless of
the shape of the perception kernel (see Appendix
\ref{Q-property}). Therefore, in a sufficiently large opinion space,
the low-frequency perturbations always destabilize the homogeneous
distribution eventually if and only if
\begin{align}
1 &> \frac{d}{c P_h}, \quad \text{or} \quad c P_h > d . \label{eq:condition2}
\end{align}
which was also confirmed through numerical simulations (see Appendix \ref{cPh-d-threshold}).

\subsection{Numerical study on the disagreement between groups}

Our main interest in this study is in the characteristic value of
$\omega$ for instability. This is because the spatial period of
perturbation, $L = 2\pi / \omega$, determines how far away the islands
of opinions are separated from each other in the opinion space, which
indicates the extent of disagreement between groups.

\begin{figure*}[htbp]
\centering
\includegraphics[width=\textwidth]{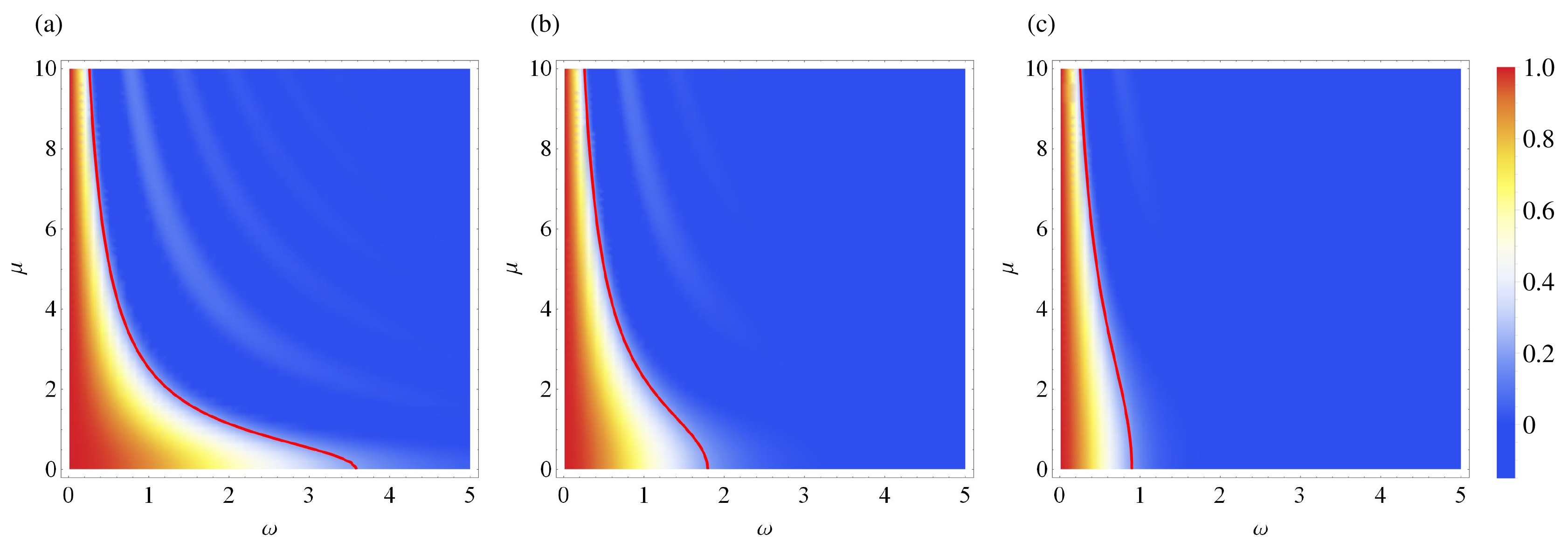}
\caption{Heatmaps showing the value of $Q(\omega)$ as a function of
  $\omega$ (horizontal axis), $\mu$ (vertical axis) and $\sigma$
  (varied in three panels; (a) $\sigma = 0.5$, (b) $\sigma = 1$, (c) $\sigma = 2$). Red curves show contours that correspond
  to $Q(\omega) = \frac{d}{c P_h} = 0.2$, the critical threshold under
  the parameter setting used for numerical simulations in this paper.}
\label{fig:omega-mu}
\end{figure*}

We numerically calculated the values of $Q(\omega)$ while varying
$\sigma$ and $\mu$. Results are shown in Fig.~\ref{fig:omega-mu}, in
which warmer colors indicate spatial frequencies $\omega$ that are
more likely to destabilize the homogeneous population distribution
(depending on the value of $\frac{d}{c P_h}$ as discussed
above). These plots show that, the greater $\sigma$ and/or $\mu$ are,
the more concentrated on low-frequency regions the unstable
perturbations are, which correspond to greater distances between
spontaneously formed groups.

\begin{figure*}[htbp]
\centering
\includegraphics[width=\textwidth]{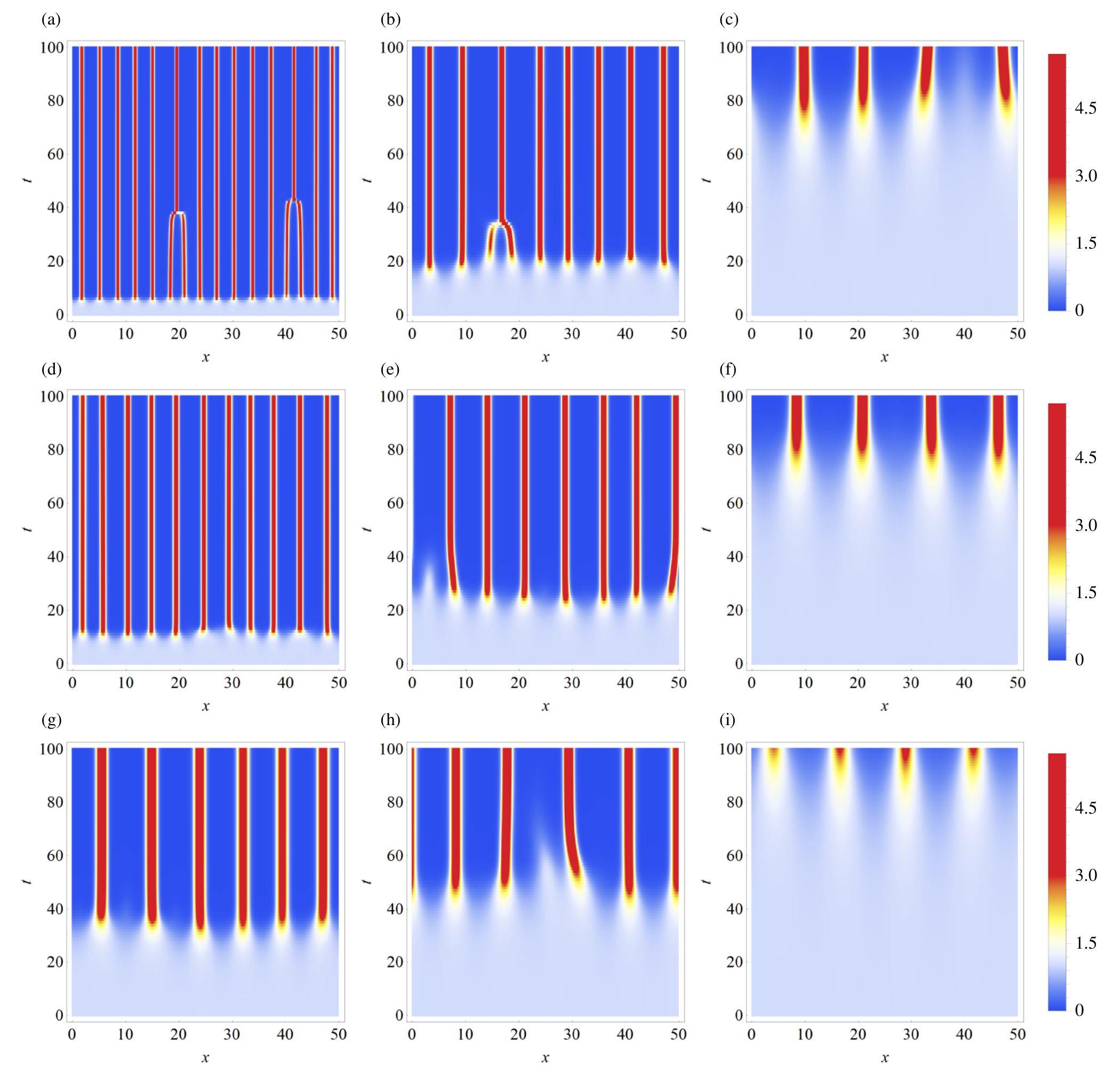}
\caption{Numerical simulation results of the population dynamics of
  the proposed model visualized in space ($x$: horizontal axis) and
  time ($t$: vertical axis, going from bottom to top). Colors
  represent population density (blue = 0, cold = low, warm = high, red
  = 3 or above). $P_h = 1$, $c = 1$, and $d = 0.2$. See the Methods section for
  details of numerical integration. Results with several different
  values of $\sigma$ and $\mu$ are shown in this figure 
  (left column (a, d, g): $\sigma=0.5$, center column (b, e, h): $\sigma=1.0$, right column (c, f, i): $\sigma=2.0$; top row (a, b, c): $\mu=0.5$, middle row (d, e, f): $\mu=1.0$, bottom row (g, h, i): $\mu=2.0$).
  The average
  distance $L$ between population peaks at
  $t = 100$ was as follows: 
  (a) ($(\sigma,\mu)=(0.5, 0.5)$): $L=3.57143$;
  (b) ($(\sigma,\mu)=(1.0, 0.5)$): $L=6.25$;
  (c) ($(\sigma,\mu)=(2.0, 0.5)$): $L=12.5$;
  (d) ($(\sigma,\mu)=(0.5, 1.0)$): $L=4.54545$; 
  (e) ($(\sigma,\mu)=(1.0, 1.0)$): $L=7.14286$; 
  (f) ($(\sigma,\mu)=(2.0, 1.0)$): $L=12.5$;
  (g) ($(\sigma,\mu)=(0.5, 2.0)$): $L=8.33333$; 
  (h) ($(\sigma,\mu)=(1.0, 2.0)$): $L=10.$;
  (i) ($(\sigma,\mu)=(2.0, 2.0)$): $L=12.5$.
}
\label{fig:patterns}
\end{figure*}

Figure \ref{fig:patterns} shows actual numerical simulation results in a space-time plot for several values of $\sigma$ and
$\mu$. Periodic boundary condition was used to avoid potential artifacts arising from cut-off spatial boundaries \footnote{Periodic boundary condition is chosen here only for numerical purposes. It would not be deemed a good assumption if this model were about polarization between two extremes. However, this is not a critical issue because the objective of this study is to characterize not such polarization but disagreement between opinionated groups, and also even the concept of polarization itself can be much more nuanced than simple binary division \cite{bramson2016disambiguation,bramson2017understanding}.}. The population distribution initially remain more or less
homogeneous for a certain period of time, but then distinct peaks
(groups) quickly form. Once established, those groups become stable
and remain unchanged for a substantially long period of time. 
Interestingly, the intervals between those peaks are longer for
greater values of $\sigma$ and/or $\mu$, which can be interpreted in
that the disagreement among those established groups becomes more
intense as people's information gathering ability is enhanced. It is
also noticeable that the effects of $\sigma$ and $\mu$ are slightly
different on the group formation process.

\begin{figure}[htbp]
\centering
\includegraphics[width=\columnwidth]{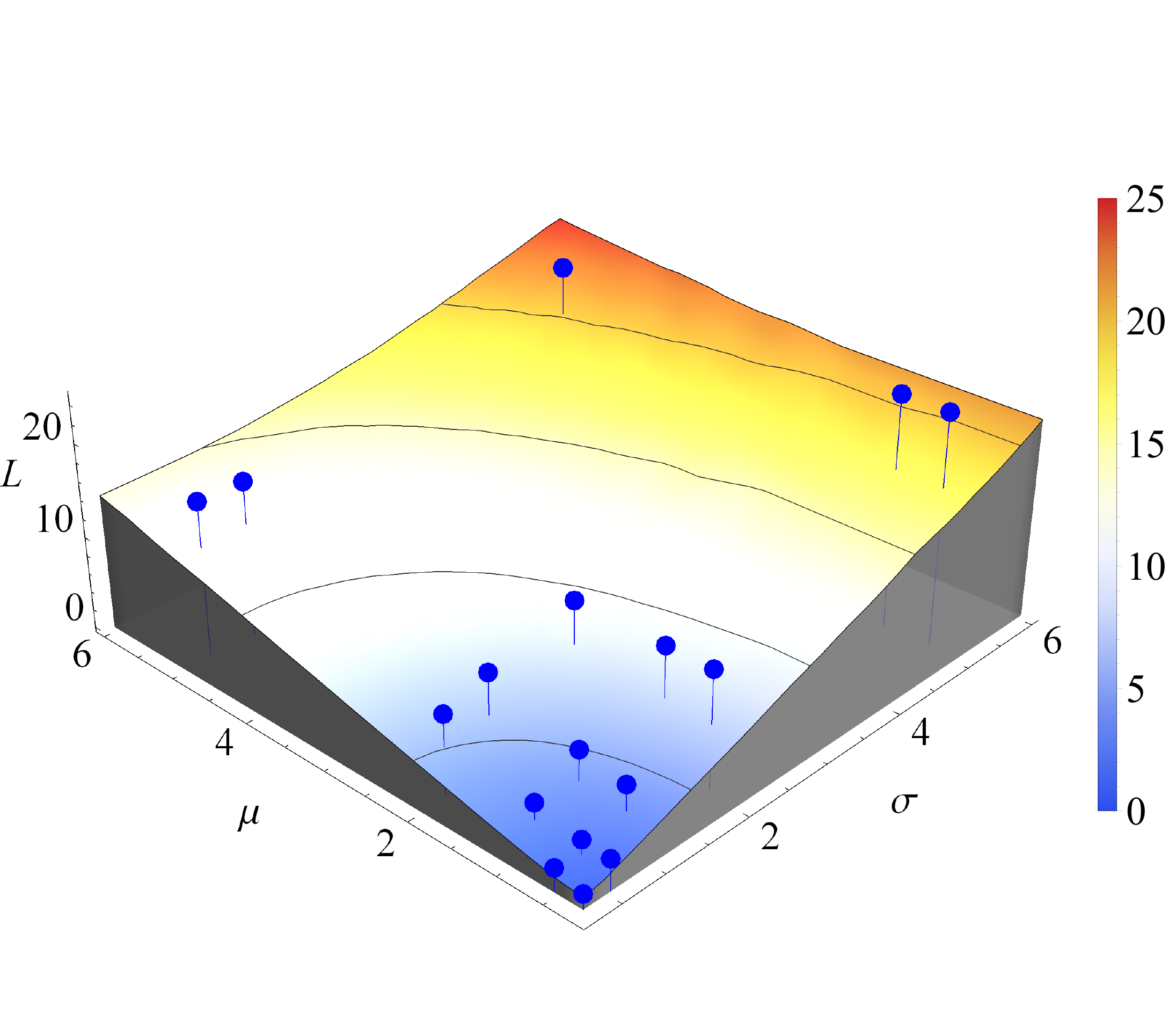}
\caption{Characteristic distance between population peaks ($L$)
  visualized as a function of $\sigma$ and $\mu$. $P_h = 1$, $c = 1$,
  $d = 0.2$. The surface plot shows a critical lower bound $L_c$ below
  which such perturbations would not grow. The lower bound $L_c$ was
  obtained as $L_c = \frac{2 \pi}{\omega_c}$, where $\omega_c$ is a
  numerically obtained critical value of $\omega$ that satisfies
  $Q(\omega) = \frac{d}{c P_h}$. The blue dots show actual peak
  distances measured in numerical simulations (e.g.,
  Fig.~\ref{fig:patterns}), which are all above the surface. This
  confirms the validity of the analytical prediction that enhanced
  information gathering ability (increased $\sigma$ and/or $\mu$)
  always results in greater distance between population peaks.}
\label{fig:sigma-mu-L}
\end{figure}

Figure \ref{fig:sigma-mu-L} summarizes these results in a single plot
of the characteristic inter-peak distance $L$ as a function of
$\sigma$ and $\mu$ for $P_h = 1$, $c = 1$ and $d = 0.2$. The surface
plot shows a numerically obtained lower bound of $L$ based on the
analysis (Eq.~(\ref{eq:condition})). Our analysis predicts that
perturbations with a characteristic length below this surface would
not grow. The peak distances obtained from numerical simulations (blue
dots in Fig.~\ref{fig:sigma-mu-L}) are all above this surface, which
confirms that our analysis was valid. It is seen in the figure that
the characteristic inter-peak distance grows almost linearly with
$\sigma$ and $\mu$, with a mild nonlinear interaction between them.

\subsection{Temporal change of information gathering ability}

This model allows us to test more dynamic scenarios in which people's
information gathering ability changes over time. Investigation of such
hypothetical scenarios can provide us with valuable insight on
potential interventions and possible societal responses to them. We
test two hypothetical dynamic scenarios below.

\begin{figure*}[htbp]
\centering
\includegraphics[width=0.7\textwidth]{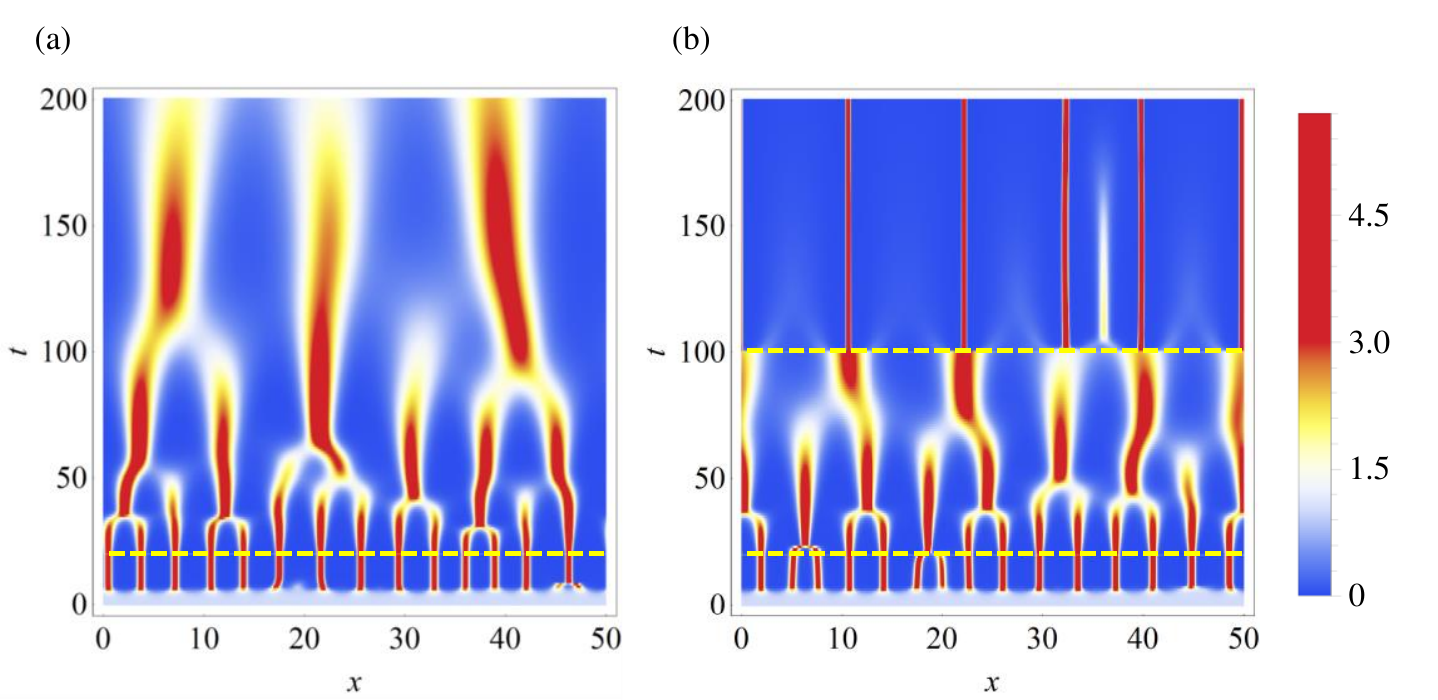}
\caption{Numerical simulation results of dynamic scenarios in which
  the values of $\sigma$ and $\mu$ are varied over time. The
  simulation method, initial condition, and meaning of colors are all
  the same as in Fig.~\ref{fig:patterns}. Simulations were conducted
  until $t=200$ in these plots. Yellow dashed lines represent key
  time points. (a) Initially $\sigma = \mu = 0.5$, but after
  $t=20$ those parameters are linearly increased with time up to $5.0$
  by $t = 200$. This scenario imitates gradual advance of information
  communication technology. As $\sigma$ and $\mu$ increase, existing
  groups tend to merge to form larger, more distant (= disagreeing)
  groups over time. (b) Conditions are the same as in (a), except
  that $\sigma$ and $\mu$ are reset to their initial value $0.5$ at $t
  = 100$ and remain constant thereafter. This scenario imitates
  external intervention to reduce people's information gathering
  ability, but it fails to diffuse the already established groups.}
\label{fig:dynamic}
\end{figure*}

The first scenario assumes that people's information gathering ability
gradually increases over time. This can be simulated by increasing
values of $\sigma$ and $\mu$ in the course of a numerical
simulation. This scenario imitates gradual advancement of information
communication technology, by which people gain access to a broader
range of opinions (by greater $\sigma$) and pay greater attention to
opinions distant from their own (by greater $\mu$). Figure
\ref{fig:dynamic}a shows an illustrative example of this scenario, in
which both $\sigma$ and $\mu$, initially set to $0.5$, begin to
increase linearly with time at $t = 20$, up to $5.0$ by the end of the
simulation at $t=200$. The smaller groups existing at the beginning
gradually merge to form larger, more distant (more disagreeing) groups
as $\sigma$ and $\mu$ increase. By the end of this particular
simulation, more than a dozen of initial small groups are integrated
into just three large groups.

The second scenario models an attempt of external intervention to the
population behavior observed in the first scenario. Specifically, we
test what would happen if people's information gathering ability were
suppressed in the middle of the first scenario with an intention to
dissolve the emerging larger groups. This scenario can be considered a
simulation of government regulation or other forms of exogenous
control, which can be simulated by lowering the values of $\sigma$ and
$\mu$ after a certain period of group formation. Figure
\ref{fig:dynamic}b shows an example of this scenario, which proceeds
the exact same way as in Fig.~\ref{fig:dynamic}a for the first half
but then $\sigma$ and $\mu$ are suddenly reset to their initial value
$0.5$ at $t=100$ and remain constant thereafter. Interestingly, the
large groups that are already established by the time of the
intervention never become diffused, but to the contrary, they become
more concentrated and more stable than before the intervention. This
is because, unlike in other spatial biological/ecological models that
have parameter-dependent characteristic scales of patterns
\cite{turing1952chemical,pearson1993complex,sayama2002spontaneous,kondo2010reaction},
groups are generally stable in auto-aggregation models like ours once
they are established, regardless of parameter values of aggregation
behavior. They may be absorbed into other groups or destroyed by
sufficiently strong diffusion, but it is extremely difficult for them
to disintegrate into smaller groups. This result implies that
suppressing people's information gathering may not work as a means to
dissolve those large groups with conflicting opinions, if they are
already established.

\section{Discussions and Conclusions}

In this study, we proposed a PDE-based mathematical model of opinion
dynamics in a continuous opinion space and studied its dynamics using
both analytical and numerical means. Contributions of this work can be
summarized in the following four points.

First, we presented an unconventional perspective to consider growing
disagreement and conflicts in society the result of spontaneous
pattern formation in an opinion space, in which the characteristic
distance between population peaks represents how severe the
disagreement is among distinct groups. This perspective allows us to
understand intensifying disagreement as a system-level global property
rather than a consequence caused by specific individuals or events to
blame. Second, we proposed a generalized non-local spatial gradient
and used it as a mathematical representation of enhanced information
gathering ability of people. This enabled us to explore different
shapes of perception kernels and also facilitated the linear stability
analysis of the model. Third, we obtained several key analytical
results on the general threshold between pattern-forming and
non-pattern-forming regimes, as well as the effects of the parameters
of information gathering behavior on the distance between resulting
groups, which were confirmed by numerical simulations. The result
clearly showed that the distance among groups became greater as
people's information gathering ability was enhanced. Finally, we
tested a few dynamic scenarios that produced relevant implications of
increasing information communication technology for social dynamics
and also some insight into the (lack of) effectiveness of external
interventions.

Our results are generally in agreement with the now commonly made
claim that the rapid development of the Internet, social media, smart
phones, and other personal information communication technologies have
contributed to the increase of societal conflicts and ideological
escalation. Our information gathering ability today is nothing
comparable even to that of twenty years ago, and such a rapid change
of our ``perception kernel'' may have already been producing emergent
macroscopic social patterns (like those shown in
Fig.~\ref{fig:dynamic}) that go beyond any single individual's
intention.

The results of the last scenario simulations illustrate challenging
aspects of the observed opinion dynamics. As the perception kernel
becomes wider and/or more focused on distant opinions, groups tend to
merge hierarchically to eventually form a small number of large groups
that are in significant disagreement from each other. Once they have
formed, reducing the perception kernel would have no effect on their
existence, but rather, it helps those groups more crystallized. This
leads us to questioning whether there are ways to remedy disagreement
between those large groups and let them gracefully revert back to
smaller groups with more distributed, more diversified opinions.

Our model suggests that, at least mathematically, several different
options exist for addressing this question. The first option is to
increase the random diffusion rate $d$ or decrease the active
migration rate $c$ so that Eq.~(\ref{eq:condition}) no longer
holds. The second option is to reduce $\mu$ all the way to a negative
value so that people seek originality rather than social conformity,
changing the dynamics of the model from auto aggregation to auto
avoidance. These two options are essentially suggesting to alter
people's behavior, but it is not obvious how one could achieve such
global behavioral changes in reality (some well-designed educational
initiatives might help). The third option is to increase $\sigma$ and
$\mu$ to extremely large values so that the boundaries of groups
become more gradual and less defined (a sign of this phenomenon is
seen near the end of the simulation in Fig.~\ref{fig:dynamic}a). The
last option suggests to {\em promote}, rather than suppress, people's
information gathering, but it would then bring another problem that
people's cognitive ability would be too limited to process the massive
amount of information collected. None of these options is
problem-free, but they may still suggest directions of potential
solutions to be explored further.

We conclude this paper by pointing out several limitations of the
study and identify future research questions. The model discussed in
this study is still quite limited in both mathematical and practical
aspects. Mathematically, we used only one form of the perception
kernel to facilitate parameterized representation of information
gathering behavior, but there should be many other functional forms
that are plausible as a model of human information gathering
behaviors. For example, it was recently reported that exposure to distant opinions may have a repulsion effect \cite{bail2018exposure}, which was not considered in the present study but could be incorporated by revising the shape and sign of the perception kernel. Exploring different forms of the perception kernel and
studying their effects on the resulting opinion dynamics would likely
produce more comprehensive understanding of this model. We also used
only one boundary condition (periodic) in all of the numerical
simulations presented, but the interactions of self-organizing
patterns with non-trivial boundaries (i.e., structure of possibilities
in the idea space) are another area that warrants further systematic
study.

In terms of practical aspects, it is with no doubt that our model
oversimplified the complexity of real human social dynamics. We
assumed only one-dimensional continuous opinion space, but opinions
and ideas can be multi-dimensional. While we expect that the essential
conclusions obtained from the stability analysis would still hold in
multidimensional opinion space, details of implications would likely
be influenced by the number of dimensions. In addition, this study did not explicitly consider non-trivial social network structure of opinion exchange. The
structure of society is implicit in this model, represented indirectly
by the perception kernel (which allows individuals whose opinion states are close to each other to interact). In other
words, the proposed model assumes that social connectivity is
dynamically correlated with the proximity of opinions of individuals
in the opinion space. Assumptions of nontrivial social network
structure with features such as heterogeneous degrees and communities are orthogonal to opinion dynamics, which can be incorporated
into the model but is beyond the scope of the present study.

Furthermore, this study
did not consider behavioral diversity of individuals within the
population at all. Such an assumption of homogeneous attributes shared
among all components is still a common practice used in many
theoretical studies of social dynamics, yet we have recently shown in
a separate study that having even a simplest kind of individual
diversity can greatly influence macroscopic behaviors of social
systems \cite{sayama2020beyond}. Introducing behavioral diversity to
the model may produce fundamentally different outcomes and
implications. Finally, the proposed model has yet to be validated in
comparison with quantitative real-world data. Eq.\ (\ref{eq:condition2}) showed a unique dimensionless quantity $\frac{d}{c P_h}$ and its critical threshold 1. This would allow at least for empirical testing of the global pattern formation condition using real-world 
data, regardless of specific choices of measurement units. Meanwhile, it might be difficult to
obtain a large-scale socio-behavioral data that could be directly used
to test the effects of the perception kernel's shape, and therefore, we anticipate the full model validation
to be done through multiple hypothesis generation and testing.

\section*{Methods}

Numerical integration of the PDEs was conducted in a discretized
space-time with spatial interval $\Delta x = 0.05$ and temporal
interval $\Delta t = 0.001$ using a simple Euler-forward numerical
integration method. The initial condition was a homogeneous population
at $P_h = 1$ everywhere in a spatial domain $[0, 50]$, with small
random perturbation (a random number sampled from a uniform
distribution $[-0.02, 0.02]$) added to each discrete spatial
location. The boundary condition was set to be periodic. 

The numerical solver was implemented by the author in Julia 1.3.0, whose source
code is available upon request. Analysis and visualization of the simulation results were conducted using Wolfram Research Mathematica 12.0.0.

\section*{Acknowledgments}

This work was supported by JSPS KAKENHI Grant Number 19K21571 and the Visitors Program of the Max Planck Institute for the Physics of Complex Systems.

H.S.\ designed the research, developed the model, conducted mathematical analysis, wrote computer simulation codes, conducted numerical experiments, analyzed the results, and wrote the manuscript.

The author does not have any competing interests.

All correspondence related to this study should be addressed to H.S. (sayama@binghamton.edu).

The data that support the findings of this study (source codes for numerical simulations, mathematical analysis, and visualization, as well as raw data for figures) are available from the author upon request.

\bibliographystyle{unsrt}
\bibliography{PDE-based-model-rev}

\newpage
\appendix

\section{Relationship between local and non-local gradients}
\label{convergence}

Here we show that the non-local gradient $G(P)$ introduced in this study converges to a simple local gradient (derivative) $P'$ in the limit of $\mu \to 0^+$ and $\sigma \to 0^+$, as follows:
\begin{widetext}
\begin{align}
\lim_{\mu, \sigma \to 0^+} G(P) 
&=\lim_{\mu \to 0^+} \frac{1}{2 \mu}
\lim_{\sigma \to 0^+} \int_{-\infty}^\infty P(x+y)
\cdot\frac{1}{\sqrt{2\pi} \sigma}
\left( e^{-\frac{1}{2} \left(\frac{y-\mu}{\sigma}\right)^2} 
- e^{-\frac{1}{2} \left(\frac{y+\mu}{\sigma}\right)^2} \right) dy \\
&= \lim_{\mu \to 0^+} \frac{P(x+\mu)-P(x-\mu)}{2 \mu} \\
&= P'(x)
\end{align}
\end{widetext}
We also confirmed this convergence numerically (results not shown).

\section{Details of linear stability analysis}
\label{stability-analysis}

We replace $P(x,t)$ in Eq.\ (\ref{eq:model}) with a homogeneous population level $P_h$ plus a sinusoidal spatial perturbation with temporally changing small amplitude $\Delta P(t)$, as follows:
\begin{align}
P(x, t) &\to P_h + \Delta P(t) \sin (\omega x + \phi)
\end{align}
Then Eq.\ (\ref{eq:model}) is rewritten as follows:
\begin{widetext}
\begin{align}
\sin (\omega x + \phi) \frac{d \Delta P}{d t} &= 
- d \omega^2 \sin (\omega x + \phi) \Delta P 
- c \frac{\partial}{\partial x} \left[
\left( P_h + \Delta P \sin (\omega x + \phi) \right)
\int_{-\infty}^\infty \left( P_h + \Delta P \sin (\omega (x+y) + \phi) \right) g(y) dy \right]
\end{align}
\end{widetext}
By ignoring the second-order term of $\Delta P$ and exploiting the fact that $g(y)$ is an odd function, this equation is linearly approximated as follows:
\begin{widetext}
\begin{align}
\sin (\omega x + \phi) \frac{d \Delta P}{d t}
&\approx - d \omega^2 \sin (\omega x + \phi) \Delta P 
- c P_h \frac{\partial}{\partial x}
\int_{-\infty}^\infty \Delta P \sin (\omega (x+y) + \phi) g(y) dy\\
&= - d \omega^2 \sin (\omega x + \phi) \Delta P 
- c \omega P_h \Delta P 
\int_{-\infty}^\infty \left[ \cos(\omega x + \phi) \cos(\omega y) - \sin(\omega x + \phi) \sin(\omega y) \right] g(y) dy\\
&= - d \omega^2 \sin (\omega x + \phi) \Delta P 
+ c \omega P_h \sin(\omega x + \phi) \Delta P 
\int_{-\infty}^\infty \sin(\omega y) g(y) dy
\end{align}
\end{widetext}
By dividing all terms by $\sin(\omega x + \phi)$ and collecting the coefficients of $\Delta P$ together, we obtain the following one-dimensional linear dynamical equation of $\Delta P$:
\begin{align}
\frac{d \Delta P}{d t} &= \left( - d \omega^2  + c \omega P_h \int_{-\infty}^\infty \sin(\omega y) g(y) dy \right) \Delta P 
\end{align}
If the coefficient inside the parentheses above is positive, the small perturbation $\sin(\omega x + \phi)$ will grow, i.e., the homogeneous population distribution will be unstable and non-homogeneous patterns (distinct groups) will form.

With $\displaystyle Q(\omega) = \int_{-\infty}^\infty \frac{\sin(\omega y)}{\omega} g(y) dy$, this condition for pattern formation is summarized as
\begin{align}
Q(\omega) &> \frac{d}{c P_h}
\end{align}
for $\omega > 0$, as described in the main text.

\section{Properties of $Q(\omega)$}
\label{Q-property}

We note that $Q(\omega)$ is, by itself, the generalized non-local gradient of $\sin(\omega x) / \omega$ around $x=0$. This indicates that the range of $Q(\omega)$ is bounded by the range of the gradients of the original function $\sin(\omega x) / \omega$, which is $\cos(\omega x)$, hence $Q(\omega) \in [-1,1]$. 

Moreover, we show that $Q(\omega)$ approaches its maximum 1 regardless of the shape of $g(y)$ in the limit of $\omega \to 0$, as follows:
\begin{align}
\lim_{\omega \to 0} Q(\omega)
&= \lim_{\omega \to 0} \int_{-\infty}^\infty \frac{\sin(\omega y)}{\omega} g(y) dy \\
&= \lim_{\omega \to 0} \int_{-\infty}^\infty y \; g(y) dy \\
& = 1
\end{align}

\section{Numerical simulations demonstrating the $c P_h > d$ instability condition}
\label{cPh-d-threshold}

\begin{figure*}[htbp]
\centering
\includegraphics[width=\textwidth]{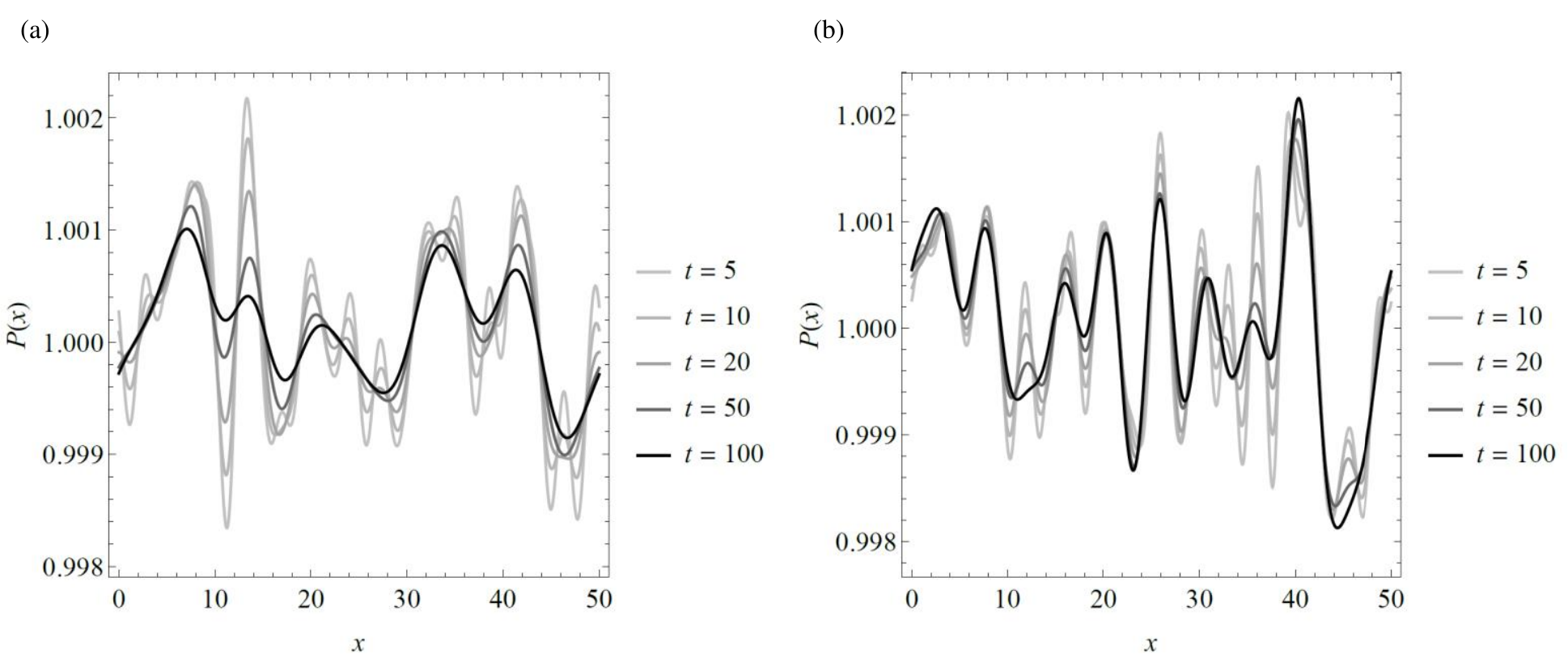}
\caption{Numerical simulation results with $c = P_h = 1$ and $\sigma =
  \mu = 0.1$. (a) Result with $d = 1.01$, which is slightly above
  $c P_h = 1$, therefore even the lowest-frequency perturbations
  gradually decay. (b) Result with $d = 0.99$, which is slightly
  below $c P_h = 1$, therefore lowest-frequency perturbations
  gradually grow and some peaks become more manifested over time. See
  the Methods section for details of numerical integration. Note that
  the vertical axis is set on a small range to show subtle difference
  between these two cases.}
\label{fig:cPh-d-threshold}
\end{figure*}

Our analysis shows that, if the opinion space is sufficiently large,
the low-frequency perturbations always destabilize the homogeneous
population distribution if and only if $c P_h > d$. This prediction
can be confirmed through numerical simulations. Illustrative cases are
shown in Fig.~\ref{fig:cPh-d-threshold}.

\end{document}